\documentclass[showpacs,aps,chaos]{revtex4}

\usepackage{amsmath}
\usepackage{amssymb}
\usepackage{graphicx}
\usepackage{dcolumn}
\usepackage{bm}

\newcommand{\ga}{\gamma}

\newcommand{\prt}{\partial}

\newcommand{\sn}{\mathrm{sn}}

\begin{document}

\title{
Self-similar wave breaking in dispersive
Korteweg-de Vries hydrodynamics}

\author{ A. M. Kamchatnov}
\affiliation{Institute of Spectroscopy,
Russian Academy of Sciences, Troitsk, Moscow, 108840, Russia}
\affiliation{Moscow Institute of Physics and Technology, Institutsky
lane 9, Dolgoprudny, Moscow region, 141700, Russia}

\date{\today}

\begin{abstract}
We discuss the problem of breaking of a nonlinear wave
in the process of its propagation into a medium at rest.
It is supposed that the profile of the wave is described at the
breaking moment by the function $(-x)^{1/n}$ ($x<0$, positive pulse)
or $-x^{1/n}$ ($x>0$, negative pulse) of the coordinate $x$.
Evolution of the wave is governed by the Korteweg-de Vries
equation resulting in formation of a dispersive shock wave.
In the positive pulse case, the dispersive shock wave forms at the
leading edge of the wave structure, and in the negative pulse case at
its rear edge. The dynamics of dispersive shock waves is described
by the Whitham modulation equations. For power law initial profiles,
this dynamics is self-similar and the solution of the Whitham equations 
is obtained in a closed form for arbitrary $n>1$.
\end{abstract}

\pacs{47.35.Jk, 47.35.Fg, 02.30.Ik }


\maketitle

{\bf Wave breaking is a universal phenomenon which takes place in evolution of
nonlinear waves in various media. In an idealized situation, when one neglects
the dissipative and dispersive effects, it leads to formation of multi-valued
solutions of the equations that describe evolution of the wave. Such a
non-physical behavior is remedied by taking into account the viscosity or
dispersion, so that the multi-valued region is replaced by a viscous or
dispersive shock wave (DSW). If the nonlinear and dispersive effects are
considered in the leading approximation, then the wave evolution is typically
governed by the Korteweg-de Vries equation. In this paper, we consider
DSWs which are generated after wave breaking of initial pulses with power-law
profiles. The advantage of this particular class of initial data is that, on one
hand, it approximates an arbitrary enough pulse near its edge and, on the other
hand, the complete solution of the Whitham equations that govern evolution of
the DSW can be obtained in a closed analytical form. This enables one to
obtain elementary formulae for such important characteristics of DSWs as velocities
ot their edges accessible for experimental measurements.
}

\section{Introduction}
\label{intro}

As is known, nonlinear waves can ``break'', or suffer from ``gradient catastrophe'',
if one does not take into account such effects as viscosity or dispersion. This
means that after some critical moment of time a formal solution of corresponding
evolution equations becomes multi-valued.
In compressible fluid dynamics, introduction into the theory of such
irreversible processes as viscosity and heat conductivity permitted one to
formulate a consistent theory of shock waves which has found many applications
(see, e.g., \cite{LL-6,CF-50}). In framework of this theory, the multi-valued
region is replaced by a narrow shock layer within which irreversible processes
dominate. In typical situations, the width of this layer has the order of
magnitude about a mean free path of molecules in the gas under consideration,
and in the macroscopic description of continuous medium dynamics such a layer
can be treated as a discontinuity in distributions of density, flow velocity
and other physical parameters of the medium.

However, there is another possible way to overcome the difficulty of
appearance of multi-valued solutions in wave dynamics. Namely,
in many physical situations the dissipative effects are negligibly small
compared with dispersion effects. In classical physics, such a
situation was considered in the theory of ``undular bores'' in shallow water waves
theory (see, e.g., \cite{bl-54}). The generality of this situation was underlined
by Sagdeev \cite{sagdeev} who indicated that in media with dispersion the macroscopic
nonlinear wave structures are generated after the wave breaking moment, and this structure
joins two flows with different values of physical parameters. At present,
such wave structures are called {\it dispersive
shock waves} and there exists considerable literature devoted to their theory
(see, e.g., the review \cite{eh-16}). Typically, a dispersive shock wave (DSW)
occupies some finite region which expands with time and consists of intensive
nonlinear wave oscillations. At one its edge the DSW can be considered as a sequence
of solitons and at another edge it degenerates into a linear harmonic wave propagating
with a certain group velocity depending on the physical situation under consideration.
The fundamental theoretical approach to the theoretical description of DSWs was
suggested by Gurevich and Pitaevskii \cite{gp-73} and it was based on
Whitham's theory of modulations \cite{whitham}. In this approach, the DSW is
represented as a modulated periodic wave of the nonlinear wave equation and
slow evolution of the modulation parameters is governed by the Whitham modulation
equations obtained by averaging the conservation laws over fast oscillations
of the physical parameters in the wave. This idea was realized by Gurevich and
Pitaevskii for waves whose evolution is described by the Korteweg-de Vries (KdV)
equation
\begin{equation}\label{eq1}
  u_t+6uu_x+u_{xxx}=0,
\end{equation}
which is derived in many typical situations in framework of the perturbation theory
in the limit of long enough wavelength and small enough (but finite) wave amplitude.

Two typical problems about wave breaking were discussed by Gurevich and Pitaevskii
in Ref.~\cite{gp-73}. First, they gave complete analytical solution of the ``dam-breaking''
problem when the initial distribution of $u_0(x)$ has a step-like form,
\begin{equation}\label{eq2}
  u_0(x)=\left\{
  \begin{array}{l}
  1,\qquad x\leq 0,\\
  0,\qquad x>0.
  \end{array}
  \right.
\end{equation}
Second, they found the main characteristics of the DSW near the generic wave breaking
point when the initial distribution can be approximated at the wave breaking moment by
a cubic parabola
\begin{equation}\label{eq3}
  u_0(x)=(-x)^{1/3}.
\end{equation}
In this case, Gurevich and Pitaevskii found velocities of the edged of the DSW and wave
amplitudes near the edges. Later, the full analytic solution of this problem was obtained
in Ref.~\cite{potemin} with the use of the inverse scattering transform method applied
to the KdV equation (see also \cite{dn-93,kamch}).

However, as is known (see, e.g., \cite{LL-6}), if the wave propagates into medium at rest,
then the wave profile at the wave breaking moment differs from (\ref{eq3}) and, depending
on the polarity of the pulse,  can be approximated either by the function
\begin{equation}\label{eq4}
  u_0(x)=\left\{
  \begin{array}{cl}
  (-x)^{1/n},\qquad &x\leq 0,\\
  0,\qquad &x>0,
  \end{array}
  \right.
\end{equation}
or by the function
\begin{equation}\label{eq4b}
  u_0(x)=\left\{
  \begin{array}{cl}
  0,\qquad &x\leq 0,\\
  -x^{1/n},\qquad &x>0,
  \end{array}
  \right.
\end{equation}
that is the wave amplitude vanishes either at the leading wave front, or at the rear edge,
according to the power law with $n>1$, and it equals to zero in the quiescent region of
the space. This problem with a positive polarity was reduced in Ref.~\cite{gkm-89} in
framework of the Gurevich-Pitaevskii
method to solution of a certain ordinary differential equation for the dependence of the
modulation parameters on the self-similar variable $z=x/t^{n/(n-1)}$, and in
Ref.~\cite{ks-90} the solution was obtained in a closed analytical form for the most
typical case with $n=2$ (see also \cite{kamch}). A similar problem with arbitrary $n$
and positive polarity was considered in Ref.~\cite{gkke-95}
in the context of supersonic flow past thin bodies in dispersive hydrodynamics.
In this paper, we will present the detailed solution of the time-dependent
evolution of the initial pulse for both cases (\ref{eq4}) and (\ref{eq4b}) and
for arbitrary value of $n>1$.

\section{Whitham modulation theory}
\label{sec:1}

At first, we shall present briefly the main necessary equations of the
Whitham theory for the KdV equation.

Periodic solution of the KdV equation can be written in the form (see, e.g., \cite{kamch})
\begin{equation}\label{eq5}
  u(x,t)=r_3+r_2-r_1-2(r_2-r_1)\sn^2(\sqrt{r_3-r_1}(x-Vt),m),
\end{equation}
where
\begin{equation}\label{eq6}
  V=2(r_1+r_2+r_3),\qquad m=\frac{r_2-r_1}{r_3-r_1},
\end{equation}
$\sn$ is the elliptic Jacobi sine function. This solution depends on three
parameters $r_1\leq r_2\leq r_3$, in terms of which we can represent the wave
velocity $V$, amplitude of oscillations $a=r_2-r_1$ in the wave and its wavelength
\begin{equation}\label{eq7}
  L=\frac{2K(m)}{\sqrt{r_3-r_1}},
\end{equation}
where $K(m)$ is the complete elliptic integral of the first kind. In the limit
$r_2\to r_3$ we get the soliton solution
\begin{equation}\label{eq8}
  u(x,t)=r_1+\frac{2(r_3-r_1)}{\cosh^2[\sqrt{r_3-r_1}(x-Vt)]},\qquad
   V=2(r_1+2r_3),
\end{equation}
and in the limit $r_2\to r_1$ the solution (\ref{eq5}) transforms into a harmonic
linear wave
\begin{equation}\label{eq9}
  u(x,t)=r_3+(r_2-r_1)\cos[2\sqrt{r_3-r_1}(x-Vt)],\qquad
  V=2(2r_1+r_3),
\end{equation}
with wavelength $L=\pi/\sqrt{r_3-r_1}$.

In a modulated wave the parameters $r_1,\,r_2,\,r_3$ become slow functions
of the space $x$ and time $t$ variables which change little in one
wavelength $L$. Their evolution obeys the Whitham equations
\begin{equation}\label{eq10}
  \frac{\prt r_i}{\prt t}+v_i(r)\frac{\prt r_i}{\prt x}=0,\quad i=1,2,3,
\end{equation}
where the Whitham velocities $v_i(r)$ can be expressed by the formula
\begin{equation}\label{eq11}
  v_i(r)=\left(1-\frac{L}{\prt L/\prt r_i}\frac{\prt}{\prt r_i}\right)V
  =V-\frac{2L}{\prt L/\prt r_i},
\end{equation}
or, after substitution of (\ref{eq7}), as
\begin{equation}\label{eq12}
  \begin{split}
  &v_1=2(r_1+r_2+r_3)+\frac{4(r_2-r_1)K(m)}{E(m)-K(m)},\\
  &v_2=2(r_1+r_2+r_3)-\frac{4(r_2-r_1)(1-m)K(m)}{E(m)-(1-m)K(m)},\\
  &v_3=2(r_1+r_2+r_3)+\frac{4(r_3-r_1)(1-m)K(m)}{E(m)},
  \end{split}
\end{equation}
where $E(m)$ is the complete elliptic integral of the second kind.
In the soliton limit $r_2\to r_3$ $(m\to1$) these velocities transform to
\begin{equation}\label{eq13}
\left.v_1\right|_{r_2=r_3}=6r_1,\qquad
  \left.v_2\right|_{r_2=r_3}=\left.v_3\right|_{r_2=r_3}=2r_1+4r_3,
\end{equation}
and in the harmonic linear limit $r_2\to r_1$ ($m\to0$) to
\begin{equation}\label{eq14}
  \left.v_1\right|_{r_2=r_1}=\left.v_2\right|_{r_2=r_1}=12r_1-6r_3,\qquad
  \left.v_3\right|_{r_2=r_1}=6r_3.
\end{equation}
Since the matrix of velocities in the system (\ref{eq10}) has a diagonal form,
the parameters $r_i$ are called Riemann invariants of the Whitham system \cite{whitham}.

\section{Dispersive shock wave formation in a positive pulse}

Now we can turn to our problem
of evolution of the wave with the initial profile (\ref{eq4}).

\begin{figure}[t]
\centerline{\includegraphics[width=13cm]{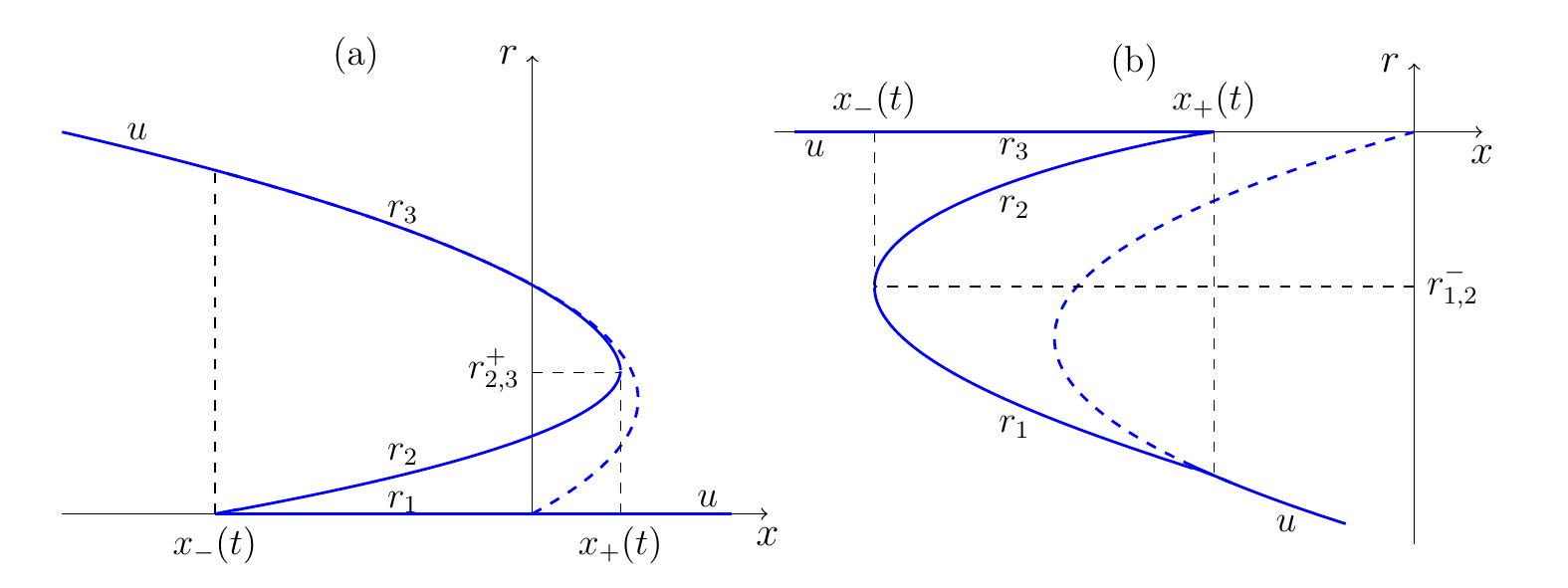}}
\caption{(a) Formal solution (\ref{eq18}) with $n=2$ is shown by thick dashed line.
It is replaced by a DSW with changing Riemann invariants $r_2$ and $r_3$
whose dependence on $x$ and $t$ is determined by the Whitham modulation equations.
(b) The Riemann invariants for a DSW generated from a negative pulse for the case $n=2$
(solid line) and a formal dispersionless solution (dashed thick line).
}
\label{fig1}
\end{figure}

In dispersionless limit, when the pulse is supposed to be a smooth enough function of $x$,
the dispersion term in Eq.~(\ref{eq1}) can be neglected and the pulse evolution
is described by the Hopf equation
\begin{equation}\label{eq15}
  u_t+6uu_x=0
\end{equation}
with well-known general solution
\begin{equation}\label{eq16}
  x-6ut=\overline{x}(u),
\end{equation}
where $\overline{x}(u)$ is a function inverse to the initial distribution $u_0(x)$.
For distributions with the form (\ref{eq4}) the formal solution (\ref{eq16}) with $\overline{x}(u)=-u^n$
has a multi-valued region which is shown in Fig.~\ref{fig1}(a) by a thick dashed line. This nonphysical behavior
is removed by means of a supposition that after wave breaking a DSW is generated which
matches at its edges $x_{\pm}(t)$ with smooth single-valued solutions of the Hopf equation in agreement with
the limiting expressions (\ref{eq13}) and (\ref{eq14})
\begin{equation}\label{eq17}
  \begin{split}
  & u=r_3\quad\text{at}\quad x=x_-\quad (m=0),\\
  & u=r_1\quad\text{at}\quad x=x_+\quad (m=1),
  \end{split}
\end{equation}
which mean that the Whitham equations for the corresponding Riemann invariants $r_3$ and $r_1$
transform to the Hopf equation (\ref{eq15}) at the edges of the DSW.
Plots of the Riemann invariants as functions of
the space coordinate $x$ at fixed time $t$ have the form shown in Fig.~\ref{fig1}(a) by a solid line
and they are similar qualitatively to the formal dispersionless solution.
At the same time, finding the laws of motion of the DSW edges $x_-(t)$ and $x_+(t)$ is part of the problem
as well as determining of the functions $r_i=r_i(x,t)$. As one can see in Fig.~\ref{fig1}(a), we assume that
in our case with the initial condition (\ref{eq4}) the Riemann invariant $r_1=0$ is constant along the DSW
in agreement with the condition $u=0$ at the soliton edge. Consequently, the equation
(\ref{eq10}) for $r_1$ is satisfied by virtue of this assumption and our wave can be called ``quasi-simple''
in accordance with definition in Ref.~\cite{gkm-89}, because only two Riemann invariants
$r_2$ and $r_3$ change along it. They are governed by the equations
(\ref{eq10}) with $v_i=r_3V_i(m)$ and
\begin{equation}\label{eq17b}
  \begin{split}
  &V_2(m)=2(1+m)-\frac{4m(1-m)K(m)}{E(m)-(1-m)K(m)},\\
  &V_3(m)=2(1+m)+\frac{4(1-m)K(m)}{E(m)},
  \end{split}
\end{equation}
where $m=r_2/r_3$.

\begin{figure}[t]
\centerline{\includegraphics[width=13cm]{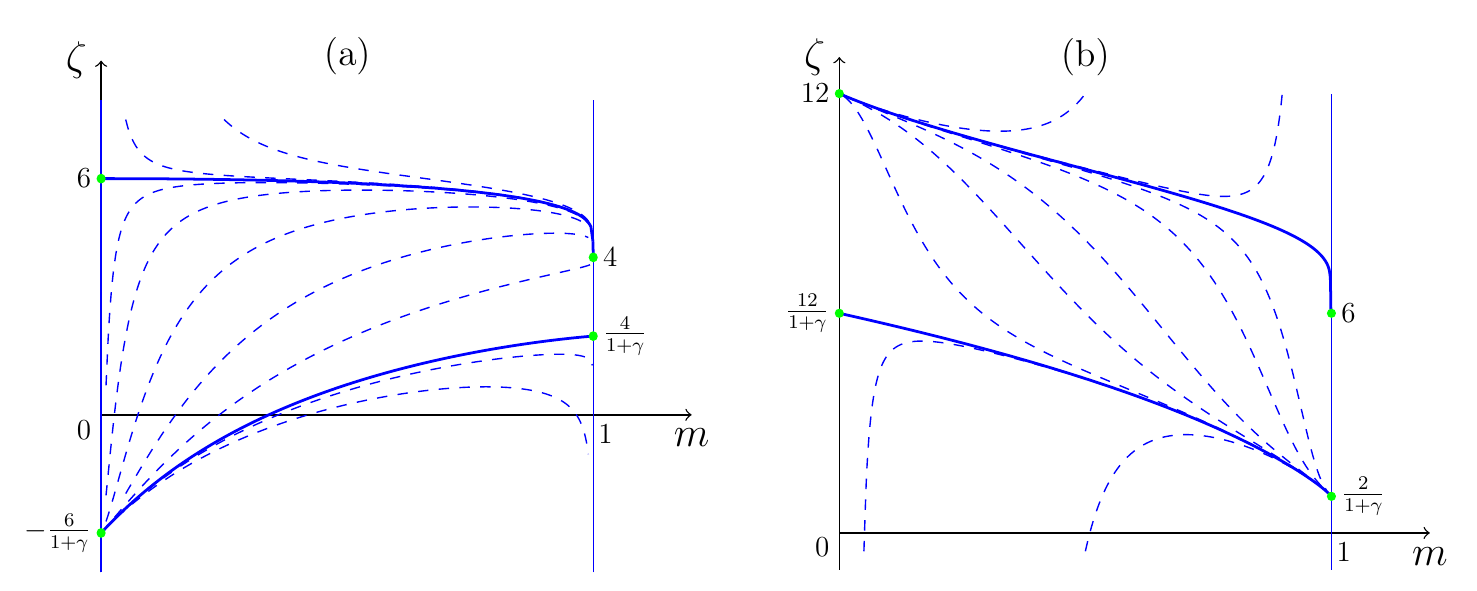}}
\caption{(a) Phase portrait of the differential equation (\ref{eq24}) in the $(m,\zeta)$ plane.
The self-similar solution corresponds to the separatrix line joining singular points
$(0,-6/(1+\gamma))$ and $(1,4/(1+\gamma))$. (b) Phase portrait of the differential equation
(\ref{eq5c}) in the $(m,\zeta)$ plane and now
the self-similar solution corresponds to the separatrix line joining singular points
$(0,12/(1+\gamma))$ and $(1,2/(1+\gamma))$.
}
\label{fig2}
\end{figure}

Now we take into account that a smooth part of the wave
\begin{equation}\label{eq18}
  x-6ut=-u^n,
\end{equation}
which is the solution of the Hopf equation, is a self-similar one: after introduction
of the variables
\begin{equation}\label{eq19}
  z=\frac{x}{t^{n/(n-1)}},\qquad u=\frac1{t^{1/(n-1)}}R(z)
\end{equation}
this solution can be rewritten as
\begin{equation}\label{eq20}
  z-6R=-R^n.
\end{equation}
The Whitham equations also admit the scaling transformation
\begin{equation}\label{eq21}
  r_i=t^{\ga}R_i(xt^{-1-\ga}),
\end{equation}
so that by taking $\ga=1/(n-1)$ we arrive at the system of equations for $R_2$ and $R_3$
as functions of the self-similar variable $z$:
\begin{equation}\label{eq22}
  \frac{dR_2}{dz}=\frac{\ga R_2}{(1+\ga)z-R_3V_2},\qquad
  \frac{dR_3}{dz}=\frac{\ga R_3}{(1+\ga)z-R_3V_3}.
\end{equation}
After transformation to the variables
\begin{equation}\label{eq23}
  m=\frac{R_2}{R_3},\qquad \zeta=\frac{z}{R_3},
\end{equation}
we obtain equation for the dependence $\zeta=\zeta(m)$:
\begin{equation}\label{eq24}
  \frac{d\zeta}{dm}=\frac{[(1+\ga)\zeta-V_2(m)][\zeta-V_3(m)]}{\ga m (V_2(m)-V_3(m))}.
\end{equation}
This equation has singular points $(0,-6/(1+\ga))$, $(0,6)$, $(1,4/(1+\ga))$, $(1,4)$
in the phase plane $(m,\zeta)$ which are shown in Fig.~\ref{fig2}(a) with account of the fact
that in our case  $n>1$ and $\ga=1/(n-1)>0$, what determines order of the points
on the line $m=1$. We are interested in the solution joining two edges of the DSW
at $m=0$ and $m=1$, hence the solution must be a separatrix in the phase plane.
At the small amplitude edge the variable $\zeta_-=z_-/R_3$ satisfies the equation
\begin{equation}\label{eq24b}
  \zeta_--6=-R_3^{n-1},
\end{equation}
and since at this point the variable $R_3$ coincides with the value of $R$ in the smooth
solution which is not equal to zero, we conclude that $\zeta(0)\neq6$, and therefore
the desired solution corresponds to the lower
separatrix. Consequently, $\zeta_-=-6/(1+\ga)=-6(1-1/n)$ and substitution of this expression
into Eq.~(\ref{eq20}) gives $R_-^{n-1}=6-\zeta_-=6(2-1/n)$, that is at the matching point
with the smooth solution at the small-amplitude edge we have
\begin{equation}\label{eq25}
  \begin{split}
   R_-=\left[6\left(2-\frac1{n}\right)\right]^{\frac1{n-1}},\qquad
   z_-=-6\left(1-\frac1n\right)\left[6\left(2-\frac1{n}\right)\right]^{\frac{1}{n-1}},
  \end{split}
\end{equation}
and this edge propagated according to the law
\begin{equation}\label{eq26}
  x_-=-6\left(1-\frac1n\right)\left[6\left(2-\frac1{n}\right)\right]^{\frac{1}{n-1}}t^{\frac{n}{n-1}}.
\end{equation}
For $n=2$ this formula reproduces the known relationship $x_-=-27t^2$ (see \cite{gkm-89,ks-90}).

For finding the general solution, we turn to the generalized hodograph method \cite{tsarev},
according to which the solution of the Whitham equations is looked for in the form
\begin{equation}\label{eq27}
  x-v_2(r)t=w_2(r),\qquad x-v_3(r)t=w_3(r).
\end{equation}
By virtue of the complete integrability of the KdV equation by the Inverse Scattering Transform
method, the functions $w_2(r),\,w_3(r)$ can be represented in the form
\begin{equation}\label{eq28}\begin{split}
  w_i(r)=\left(1-\frac{L}{\prt L/\prt r_i}\frac{\prt}{\prt r_i}\right)W=
  W-\frac12(V-v_i(r))\frac{\prt W}{\prt r_i},\qquad i=2,3,
  \end{split}
\end{equation}
similar to Eq.~(\ref{eq11}). Then, as was shown in Refs.~\cite{gke-92,wright,tian},
the function $W$ must satisfy the Euler-Poisson equation
\begin{equation}\label{eq29}
  \frac{\prt^2W}{\prt r_2\prt r_3}
  -\frac1{2(r_2-r_3)}\left(\frac{\prt W}{\prt r_2}-\frac{\prt W}{\prt r_3}\right)=0.
\end{equation}
In our self-similar case it follows from Eq.~(\ref{eq28}) that $W$ is a uniform
function of the order $n$, that is it can be represented as
\begin{equation}\label{eq30}
  W(r_2,r_3)=r_3^n\widetilde{W}(m),\qquad m=r_2/r_3.
\end{equation}
Substitution of this expression into (\ref{eq29}) yields the hypergeometric equation
for $\widetilde{W}$,
\begin{equation}\label{eq31}
  m(1-m)\frac{d^2\widetilde{W}}{dm^2}+\left[\frac12-n-\left(\frac32-n\right)m\right]
  \frac{d\widetilde{W}}{dm}+\frac{n}2\widetilde{W}=0.
\end{equation}
As is known, (see, e.g., \cite{ww}), pairs of its basis solutions can be chosen by
three different ways depending on behavior of the solution at the singular points,
and we have to choose a linear combination of any pair such, that it corresponds to
the above mentioned separatrix solution of Eq.~(\ref{eq24}). As we shall see,
this condition is fulfilled for the basis solution
\begin{equation}\label{eq32}
  \widetilde{W}(m)=C_nF(-n,1/2;1;1-m),
\end{equation}
where $F$ is a standard hypergeometric function (see, e.g., \cite{ww}) and the constant
$C_n$ is chosen according to the condition that for $m=0$ the function
$w_3$, which is determined by the formula (\ref{eq27}), matches with the smooth solution
(\ref{eq18}) of the Hopf equation, that is $w_3(0,r_3)=-r_3^n$. As a result, we obtain
\begin{equation}\label{eq33}
  C_n=-\frac{4^n[\Gamma(n+1)]^2}{\Gamma(2(n+1))}.
\end{equation}
Since the solution (\ref{eq32}) is a regular function on the closed interval $0\leq m\leq1$, it
is clear without any calculations that it corresponds to the separatrix solution, which
joins the singular points at $m=0$ and $m=1$.

Representing $w_i$ in a self-similar form
\begin{equation}\label{eq34}
  w_i=r_3^nW_i(m),\qquad i=2,3,
\end{equation}
we find with the help of Eq.~(\ref{eq28}) the expressions
\begin{equation}\label{eq35}
  \begin{split}
  W_2=&C_n\Big[F(-n,1/2;1;1-m)-\frac{n}2\left(1+m-\frac12V_2(m)\right)
   F(1-n,3/2;2;1-m)\Big],\\
  W_3=&C_n\Big[F(-n,1/2;1;1-m)-\frac{n}2\left(1+m-\frac12V_3(m)\right)\times\\
  &\times\Big(2F(-n,1/2;1;1-m)-mF(1-n,3/2;2;1-m)\Big)\Big].
  \end{split}
\end{equation}
As a result, we find the solution of our problem in a parametric form, where all
variables are expressed as functions of $m$:
\begin{equation}\label{eq36}
  z(m)=\frac{(W_2V_3-W_3V_2)(V_3-V_2)^{1/(n-1)}}{(W_2-W_3)^{n/(n-1)}},
\end{equation}
\begin{equation}\label{eq37}
\begin{split}
  & R_3(m)=\left(\frac{V_3-V_2}{W_2-W_3}\right)^{1/(n-1)},\qquad
   R_2(m)=mR_3(m),
  \end{split}
\end{equation}
\begin{equation}\label{eq38}
  \zeta(m)=\frac{W_2V_3-W_3V_2}{W_2-W_3}.
\end{equation}
The last formula gives a closed analytic expression for the solution of Eq.~(\ref{eq24}).
It is easy to see that for $m=1$ it gives $\zeta(1)=4(1-1/n)$, that is $\zeta(m)$
corresponds indeed to the lower separatrix in Fig.~\ref{fig2}(a).
Equations (\ref{eq36}) and (\ref{eq37}) allow us to obtain closed expressions for
$z_+$ and $R_3^+$ at the leading soliton edge of the DSW propagating into a quiescent
medium,
\begin{equation}\label{eq39}
  z_+=\left(1-\frac1n\right)\left\{\frac{\Gamma(2(n+1))}{n[\Gamma(n+1)]^2}\right\}^{1/(n-1)},
\end{equation}
\begin{equation}\label{eq40}
  R_3^+=\left\{\frac{\Gamma(2(n+1))}{4^{2n-1}n[\Gamma(n+1)]^2}\right\}^{1/(n-1)}.
\end{equation}
Thus, the leading edge propagates according to the law
\begin{equation}\label{eq41}
  x_+(t)=\left(1-\frac1n\right)\left\{\frac{\Gamma(2(n+1))}{n[\Gamma(n+1)]^2}\right\}^{1/(n-1)}t^{n/(n-1)}.
\end{equation}
For $n=2$ this formula gives $x_+=(15/2)t^2$ in agreement with the known result \cite{gkm-89,ks-90}.

The solution found here is correct for any value of $n>1$. It simplifies for integer $n$,
when the hypergeometric function reduces to the Jacobi polynomials
$P_n^{(1/2-n,0)}(1-2m)$, that is
\begin{equation}\label{eq42}
  W=(-1)^{n-1}r_3^n\cdot\frac{4^n[\Gamma(n+1)]^2}{\Gamma(2(n+1))}\cdot P_n^{(1/2-n,0)}(1-2m).
\end{equation}
In particular, for the case $n=2$ we find
\begin{equation}\label{eq43}
  W=-\frac{r_3^2}5\left(1+\frac25m+m^2\right)=-\frac15\left(r_2^2+\frac23r_2r_3+r_3^2\right),
\end{equation}
\begin{equation}\label{eq44}
  \zeta(m)=\frac{(1+m)(3+2m+3m^2)E(m)-(1-m)(3+14m-9m^2)K(m)}{(3+2m+3m^2)E(m)-(1-m)(3+m)K(m)},
\end{equation}
\begin{equation}\label{eq45}
  R_3(m)=\frac{15[(1+m)E(m)-(1-m)K(m)]}{(3+2m+3m^2)E(m)-(1-m)(3+m)K(m)}.
\end{equation}

We illustrate the obtained solution by two more typical examples.
For $n=3$ we get
\begin{equation}\label{eq46}
  x_-=-4\sqrt{10}\,t^{3/2},\quad x_+=\frac43\sqrt{\frac{35}3}\,t^{3/2},
\end{equation}
and for $n=3/2$ we obtain
\begin{equation}\label{eq47}
  x_-=-128\,t^{3},\quad x_+=\frac13\left(\frac{256}{9\pi}\right)^2t^{3},
\end{equation}

\begin{figure}[t]
\centerline{\includegraphics[width=13cm]{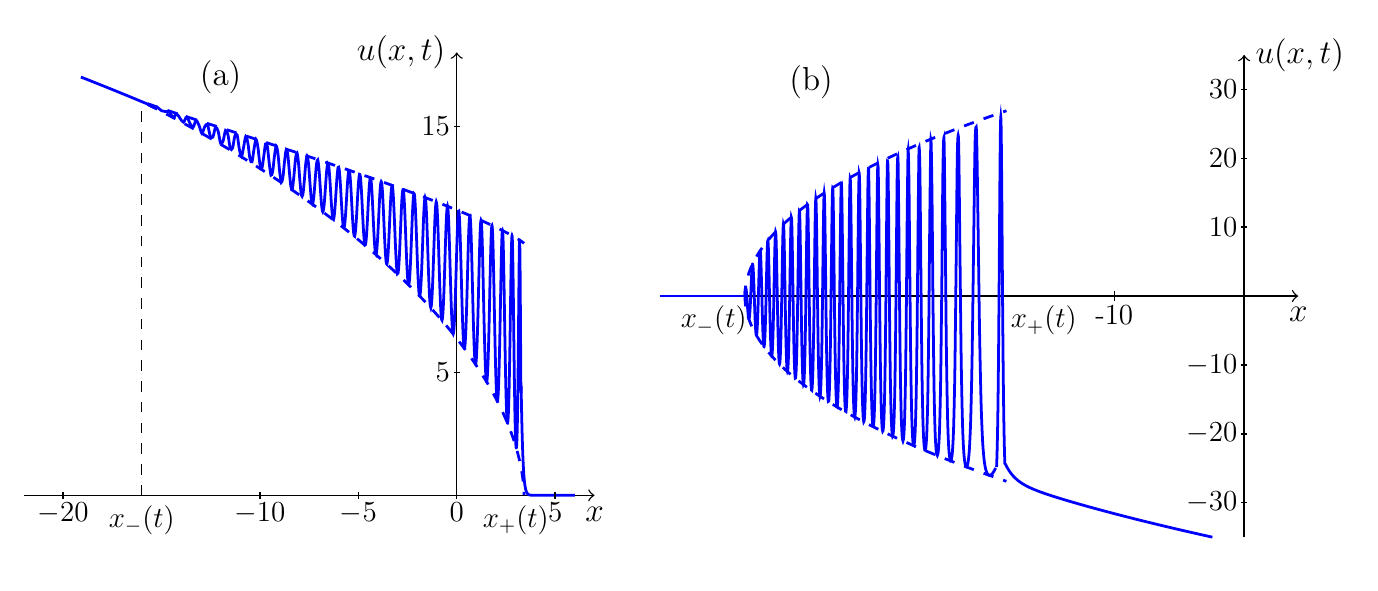}}
\caption{(a) Dispersive shock wave generated after wave breaking of
a ``positive'' pulse with the profile (\ref{eq4}) with $n=3/2$.
(b) Dispersive shock wave generated after wave breaking of
a ``negative'' pulse with the profile (\ref{eq4b}) with $n=3/2$.
}
\label{fig3}
\end{figure}

The DSW arising as a result of wave breaking is described by the formulae obtained after
substitution of the functions $r_2(x,t)$, $r_3(x,t)$, defined parametrically, into
Eq.~(\ref{eq5}). For example, the DSW for the case $n=3/2$ is shown in Fig.~\ref{fig3}(a).
Thus, the solution obtained here provides a simple enough formulae for all most
important parameters of the DSW evolved from the initial profile (\ref{eq4}).

\section{Dispersive shock wave formation in a in a negative pulse}

The theory of DSW forming from a negative pulse is very similar to that for a positive pulse,
but the details are different. Therefore we give here only a concise exposition of this theory.

In this case, the diagram of the Riemann invariants has the form depicted in Fig.~\ref{fig1}(b),
that is we have here a quasi-simple wave with $r_3=0$ and self-similar dependence of two other
Riemann invariants
\begin{equation}\label{eq1c}
  r_1=t^{\gamma}R_1(z),\qquad r_2=t^{\gamma}R_2(z),\qquad z=x/t^{1+\gamma},\qquad \gamma=1/(n-1),
\end{equation}
where $R_1$ and $R_2$ satisfy the equations
\begin{equation}\label{eq2c}
  \frac{dR_1}{dz}=\frac{\ga R_1}{(1+\ga)z-R_1V_1},\qquad
  \frac{dR_2}{dz}=\frac{\ga R_2}{(1+\ga)z-R_1V_2},
\end{equation}
and now
\begin{equation}\label{eq3c}
  \begin{split}
  &V_1(m)=2(2-m)-\frac{4mK(m)}{E(m)-K(m)},\\
  &V_2(m)=2(2-m)+\frac{4m(1-m)K(m)}{E(m)-(1-m)K(m)},
  \end{split}
\end{equation}
where $m=1-R_2/R_1$. For the variable
\begin{equation}\label{eq4c}
  \zeta=\frac{z}{R_1}
\end{equation}
we obtain the differential equation
\begin{equation}\label{eq5c}
  \frac{d\zeta}{dm}=\frac{[\zeta-V_1(m)][(1+\ga)\zeta-V_2(m)]}{\ga (1-m) (V_1(m)-V_2(m))},
\end{equation}
and the self-similar solution, that we are looking for, corresponds to the separatrix
solutions of Eq.~(\ref{eq5c}) which joins two its singular points $(0,12)$, $(0,12/(1+\ga))$, $(1,6)$,
$(1,2/(1+\ga))$. At the soliton edge of DSW we have now
\begin{equation}\label{eq6c}
  6-\zeta_+=(-R_1)^{n-1},
\end{equation}
where $R_1$ matches with the smooth solution with non equal to zero value or $R$. Hence we arrive
at the conclusion that the self-similar solution corresponds to the lower separatrix in Fig.~\ref{fig2}(b).
This gives
\begin{equation}\label{eq7c}
  \zeta_+=\frac2{1+\ga}=2\left(1-\frac1n\right),\qquad R_+=-\left(4+\frac2n\right)^{\frac1{n-1}},
  \qquad z_+=\zeta_+R_+,
\end{equation}
and, correspondingly, we find the low of motion of the leading edge of the DSW,
\begin{equation}\label{eq8c}
  x_+=-2\left(1-\frac1{n}\right)\left(4+\frac2n\right)^{\frac1{n-1}}t^{\frac{n}{n-1}}.
\end{equation}

To find the global solution of the Whitham equations including the law of motion of the small-amplitude
edge, we assume that it has the form
\begin{equation}\label{eq9с}
  x-v_1(r)t=w_1(r),\qquad x-v_2(r)t=w_2(r)
\end{equation}
with
\begin{equation}\label{eq10c}
\begin{split}
  w_i(r)=  W-\frac12(V-v_i(r))\frac{\prt W}{\prt r_i},\qquad i=1,2,
  \end{split}
\end{equation}
and then the function $W$ must satisfy the Euler-Poisson equation
\begin{equation}\label{eq11c}
  \frac{\prt^2W}{\prt r_1\prt r_2}
  -\frac1{2(r_1-r_2)}\left(\frac{\prt W}{\prt r_1}-\frac{\prt W}{\prt r_2}\right)=0.
\end{equation}
In our self-similar case we look for its solution in the form
\begin{equation}\label{eq12c}
  W(r_1,r_2)=(-r_1)^n\widetilde{W}(m),\qquad m=1-r_2/r_1.
\end{equation}
Substitution of this expression into (\ref{eq11c}) yields the hypergeometric equation
for $\widetilde{W}$,
\begin{equation}\label{eq13c}
  m(1-m)\frac{d^2\widetilde{W}}{dm^2}+\left[1-\left(\frac32-n\right)m\right]
  \frac{d\widetilde{W}}{dm}+\frac{n}2\widetilde{W}=0.
\end{equation}
The separatrix solution corresponds now to
\begin{equation}\label{eq14c}
  \widetilde{W}(m)=C_nF(-n,1/2;1;m),
\end{equation}
where the constant $C_n$ is determined by the matching condition
$x-6r_1t=(-r_1)^n$ at the leading soliton edge what gives
\begin{equation}\label{eq15c}
  C_n=\frac{4^n[\Gamma(n+1)]^2}{\Gamma(2(n+1))}.
\end{equation}
Thus, we get the solution
\begin{equation}\label{eq16c}
  w_i=(-r_1)^n W_i(m),\qquad i=1,2,
\end{equation}
where
\begin{equation}\label{eq17c}
  \begin{split}
   W_1=&C_n\Big[F(-n,1/2;1;1-m)-\frac{n}2\left(2-m-\frac12V_1(m)\right)\times\\
  &\times\Big(2F(-n,1/2;1;m)-(1-m)F(1-n,3/2;2;m)\Big)\Big],\\
   W_2=&C_n\Big[F(-n,1/2;1;m)-\frac{n}2\left(2-m-\frac12V_2(m)\right)
    F(1-n,3/2;2;m)\Big].
  \end{split}
\end{equation}
As a result, the dependence of the Riemann invariants and other functions
on the self-similar variable $z$ can be representer in a parametric form,
\begin{equation}\label{eq18c}
  z(m)=\frac{(W_2V_1-W_1V_2)(V_1-V_2)^{1/(n-1)}}{(W_1-W_2)^{n/(n-1)}},
\end{equation}
\begin{equation}\label{eq19c}
\begin{split}
  & R_1(m)=-\left(\frac{V_1-V_2}{W_1-W_2}\right)^{1/(n-1)},\qquad
   R_2(m)=(1-m)R_1(m),
  \end{split}
\end{equation}
\begin{equation}\label{eq20c}
  \zeta(m)=\frac{V_2W_1-V_1W_2}{W_1-W_2}.
\end{equation}
These expressions yield the values of the variable $R_1$ at the rear edge
\begin{equation}\label{eq21c}
  R_1^-=-\frac14\left\{\frac{3\Gamma(2n+1)}{n[\Gamma(n+1)]^2}\right\}^{1/(n-1)}
\end{equation}
and, hence, the law of motion of this edge is given by the formula
\begin{equation}\label{eq22c}
  x_-(t)=-3\left(1-\frac1n\right)\left\{\frac{3\Gamma(2n+1)}{n[\Gamma(n+1)]^2}\right\}^{1/(n-1)}t^{n/(n-1)}.
\end{equation}.

For integer values of $n$ the function $W$ reduces to a polynomial form and these
expressions can be simplified. In particular, for $n=2$ we obtain
\begin{equation}\label{eq23c}
  W=-\frac15\left(r_1^2+\frac23r_1r_2+r_2^2\right),
\end{equation}
\begin{equation}\label{eq24c}
  \zeta(m)=\frac{(2-m)(3m^2-8m+8)E(m)+2(1-m)(3m^2+8m-8)K(m)}{(3m^2-8m+8)E(m)-4(1-m)(2-m)K(m)},
\end{equation}
\begin{equation}\label{eq25c}
  R_1(m)=-\frac{15[(2-m)E(m)-2(1-m)K(m)]}{(3m^2-8m+8)E(m)-4(1-m)(2-m)K(m)},
\end{equation}
We illustrate the solution found by several examples:
\begin{equation}\label{t3-145.24}
\begin{split}
  & x_-=-\left(\frac{8}{3}\right)^3\,t^{3},\quad x_+=-\left(\frac{64}{3\pi}\right)^2t^{3}
  \quad\text{for}\quad n=3/2;\\
  & x_-=-\frac{27}2t^2,\quad x_+=-5t^2 \quad\text{for}\quad n=2;\\
  & x_-=-\frac43\sqrt{\frac{14}3}\,t^{3/2},\quad x_+=-4\sqrt{5}\,t^{3/2},
  \quad\text{for}\quad n=3;\\
  \end{split}
\end{equation}
Again the DSW arising as a result of wave breaking of a negative pulse
for the case $n=3/2$ is shown in Fig.~\ref{fig3}(b). As we see, it has a quite
different form compared with the DSW generated by a positive pulse which is
shown in Fig.~\ref{fig3}(a).

\section{Conclusion}

We have obtained the complete solution of the wave breaking problem in the KdV
equation theory for the case of power dependence of the initial positive or
negative pulse profiles on the coordinate when this problem is self-similar.
Since any initial monotonous profile can be represented as a sum of functions of the form
(\ref{eq4}), then the solution of the hodograph equations is a sum of the solutions
obtained in this paper. Therefore they provide the method of description of DSW
propagation in more general situations. Thus, the solution found can be used as 
a practical tool in comparison of experimental results with theoretical predictions
of the KdV approximation. Besides that, the laws of motion of the DSW edges can serve
as a basis for development of more general theory applicable to DSWs whose evolution is
governed by non-integrable equations (see. Refs.~\cite{el-05,kamch-18}).

\begin{acknowledgments}
I am grateful to M.~Isoard, S.~K.~Ivanov and N.~Pavloff for useful discussions.
\end{acknowledgments}

\end{document}